\documentclass{ws-procs961x669}          

\newcommand*{\FigPath}{./Figs}
\newcommand{\trho}{{\tilde \rho}}\def\cG{{\cal G}}
\def\r{\rho}\def\d{\delta}\def\bfb{{\bf b}}\def\l{\lambda}
\newcommand{\bea}{\begin{eqnarray}}
\newcommand{\eea}{\end{eqnarray}}
 \newcommand{\bfR}{{\bf R}}\newcommand{\tr}{\mathrm{tr\, }}\newcommand{\bcl}[3]{b^\dagger_{#1,#2,#3}}\newcommand{\bal}[3]{b_{#1,#2,#3}}\newcommand{\bra}[1]{\Bigl<#1\Bigl|}
\newcommand{\ket}[1]{\Bigr|#1\Bigr>}\def\bfk{{\bf k}}

\def\D{\Delta}

\begin{document}
\title{Color Correlations in the  Proton}

\author{Gerald.~A.~Miller$^*$ }

\address{Physics  Department, University of Washington,\\
Seattle, WA 98195-1560, USA\\
$^*$E-mail:miller@uw.edu \\
www.washington.edu}

\begin{abstract}
A general QCD light front formalism to compute many-body color charge correlation functions due to quarks  in the proton was constructed~\cite{Dumitru:2018vpr}. These enable new studies of color charge distributions in the nucleon. The analogies between such correlation functions   to electric and magnetization charge densities in the proton and  also to nucleon-nucleon correlations in the nucleus are discussed.  Including  the color charge correlations leads naturally to the removal of infrared divergences that occurs in two- and three-gluon exchange interactions in $q\bar q$-proton scattering.  Extensions to include  gluonic color charge correlations are  discussed.  \end{abstract}


\bodymatter

\section{Introduction}\label{aba:sec1}
This talk is concerned with a basic question:  Where is the color charge located in a nucleon, or in a
nucleus?
A related question: Is the color charge distribution the same as the charge distribution? We know that this cannot be the case because the integral of the color charge density must vanish because of color neutrality.  
Instead one must be concerned with matrix elements of powers of the color charge density operators such as $\rho^2$ and $\rho^3$.  Thus moments  must be constructed and understood.  This talk is a new way of looking at nucleon and nuclear structure. The formalism~\cite{Dumitru:2018vpr} can be thought of as an extension  of the 
  Mclerran-Venugopalan (MV) model for relativistic heavy ion physics~\cite{McLerran:1993ni,McLerran:1993ka,McLerran:1994vd} to nucleon structure. 

\section{Electromagnetic Charge Densities and Correlations}
I set the stage by discussing electromagnetic charge densities within the light-front formalism~\cite{Miller:2007uy,Carlson:2007xd,Miller:2009qu,Miller:2010nz}. 
The electromagnetic current density is given by $J^\mu=\sum_q e_q \bar q \gamma^\mu q$ where $q$ represents the quark flavor. One forms the current density from the $J^+$ operator acting in an eigenstate of transverse position $\bfR$:
\bea
\rho_\infty(x^-,\bfb)=\langle p^+,\bfR,\l|\sum_qe_qq_+^\dagger(x^-,\bfb)q_+(x^-,\bfb)|p^+,\bfR,\l\rangle,
\eea
where the longitudinal (transverse) spatial coordinates are $x^-\propto(t-z)$ (\bfb), and $q_+\propto\gamma^0\gamma^+q$ is the independent quark-field operator.   The value of $p^+$ must be very large because of the inherent sum over all of the proton's transverse momenta.  The transverse density can then be constructed  from the Dirac form factor $F_1$ as:
\bea \r(b)=\int dx^-\r_\infty(x^-,\bfb)=\int {QdQ\over 2\pi}F_1(Q^2)J_0(Qb).\eea
  Results are shown in the cited references and in the talk   posted on the INT website  http://www.int.washington.edu/talks/WorkShops/{\rm int}$_{-}18_{-}3$.
  
  Another question can be asked: given a $u$-quark at a position $(x^-,\bfb)$, what is the probability $P(\D x^-,\D\bfb)$ that a $d$-quark is a distance $(\D x^-,\D\bfb)$ away? This given by
  \bea P(\Delta x^-, \Delta \bfb)=\langle p^+{\bf R},\lambda|\int dx^-d^2\bfb \rho_u (x^-,\bfb) \rho_d (x^-+\Delta x^-,\bfb+\Delta \bfb)|p^+{\bf R}\lambda\rangle \eea
where
$\rho_q(x^-,\bfb)\equiv e_q q^\dagger_+(x^-,\bfb) q_+(x^-,r_\perp)$.  This represents a correlation function, the matrix-element of a two-quark operator that  enters in the evaluation of the two-photon-exchange matrix element. The matrix element of Eq.(3) is analogous to the short-ranged nucleon-nucleon correlations that are now under investigation. See {\it e.g.}  \cite{Hen:2016kwk,Cruz-Torres:2017sjy}.  Interactions between high energy particles and the proton are governed by two-gluon exchanges, so that probing quark-quark correlations might be easier than with two-photon exchanges.
\section{Color Charge  Density Operator }
The color charge density operator is given by 
\bea\rho^{ a} (x)=
\bar\psi_{i,f}(x)\, \gamma^+ \,\psi_{j,f}(x) (t^a)_{ij}+\,\rm{ gluon\,terms},\eea
where $a$ is the color index and $t^a$, $a=1-8$
are the generators of the fundamental representation of color-SU(3)
normalized as $\tr t^a t^b=\delta^{ab}/2$. 
The interesting matrix elements  in the proton are
$\langle \,{\rm }|\rho^a(x)|{\rm } \rangle, \langle \,{\rm }|\rho^a(x)\rho^b(y)|{\rm } \rangle, \langle \,{\rm }|\rho^a(x)\rho^b(y)\rho^c(y)|{\rm } \rangle$. The use of moments of the color charge density originated in the MV model, which shows how 
observables may be computed in terms of moments of $\rho^a$.

A few details are presented. When using light-front dynamics, quark-fields  at $x^+=0$  are  expanded in terms of    creation  and  destruction operators. The present evaluation ignores the presence of anti-quarks in the proton. Then using $r=(x^-,\bfb)$  
\bea
\rho^a(r) = 2 P^+\sum_{\lambda,\lambda'} \int {d x_q d^2q \over  {16\pi^3\sqrt{x_q}}}
\, \bcl{q}{i}{\lambda} e^{i q\cdot r}
\int {d x_p d^2p \over {16\pi^3\sqrt{x_p}}}\, \bal{p}{j}{\lambda'}
 e^{-i p\cdot r} \, (t^a)_{ij}\, \delta_{\lambda \lambda'}.\eea 
It is interesting to evaluate this operator in  the infinite momentum frame (IMF), $P^+\to\infty$. Then, the expression 
$ \lim_{P^+\to\infty} P^+ e^{i(x_q-x_p)P^+ r^-}$ appears. Taking the limit carefully~\cite{Dumitru:2018vpr}, we found that $\r^a$ contains a factor $2\pi \delta(x_p-x_q)\delta(r^-) $. Thus in the IMF the color-charge-density operator takes on the characteristics of a very thin disk.
In particular, the 
color charge per unit area is given by matrix elements of 
\bea
\rho^a(x^-,\bfb) = \d(x^-)\int \frac{d^2k}{(2\pi)^2}
\, e^{i\bfk\cdot\bfb}
 \int_0^\infty {dq^+\over q^+}\int {d^2q\over 16\pi^3}\sum_\lambda
\bcl{x_q,\vec q-\vec k}{i}{\lambda} \, \bal{x_q,\vec q}{j}{\lambda} \, (t^a)_{ij} .
\eea
The two-dimensional Fourier transform $\rho^a(x^-,\vec k_\perp) $ is used also.
 The notation \bea
\langle {\bf O}\,\rangle_{K_\perp} = \frac{\bra{P^+,\vec K_\perp} \, {\bf O}\, \ket{P^+, \vec P_\perp=0}}{\langle K|P\rangle}
\eea is used in the following.

\section{Color Charge Correlations}
Evaluations were made using light front wave functions for a 3-quark Fock state \cite{Dumitru:2018vpr}. The first result is that 
$ \langle \rho^a(x^-,\bfb)\rangle_{K_\perp}=0$ as expected for a color singlet. But the obvious result raises a question. Consider the matrix element of $\r^a$ for a 
Fock space component of proton: $|3q,{\rm G}\rangle$. Would the matrix element still vanish? The sum of quark and gluon densities must vanish, so any non-zero contribution to the color charge density from the quarks would be cancelled by the contribution from the gluons. That it is highly likely that  $\langle3q, {\rm G}| \rho^a(x^-,\bfb)|3q, {\rm G}\rangle$ would not vanish can be seen immediately  by considering color SU(2). In this case, the situation is analogous to that of the nucleon's pion cloud, which gives the neutron a non-vanishing charge density even though the total charge is zero \cite{Miller:2002ig}.
 
 The next step was to evaluate the two-quark color charge density:
$
\langle\, \tilde \rho^a(\vec K_\perp -\vec k_\perp) \, \trho^b(\vec k_\perp) \,\rangle_{K_\perp} =  \frac{1}{2}\,\delta^{ab}( \cG_1(\vec K_\perp)- \cG_2(\vec k,\vec K_\perp))\equiv \frac{1}{2}\,\delta^{ab}\cG(\vec k_\perp,K_\perp)$. The first term occurs when both density operators act on the same quark, and the second occurs when the action is on two different quarks. The limit of forward scattering is $\vec K_\perp=0$ and then   
$\cG(\vec k,0)=1-\cG_2(\vec k,0) $. Note that  taking $k_\perp=0$ yields $\cG(0,0) =0$, a consequence of color neutrality,   necessary to suppress   infrared divergences that would come from gluon propagators. 

A simple three-quark wave function \cite{Frank:1995pv} was used to evaluate $\cG(\vec k,0)$. The result is shown in Fig.~1.
 \begin{figure}
\includegraphics[width=3.in]{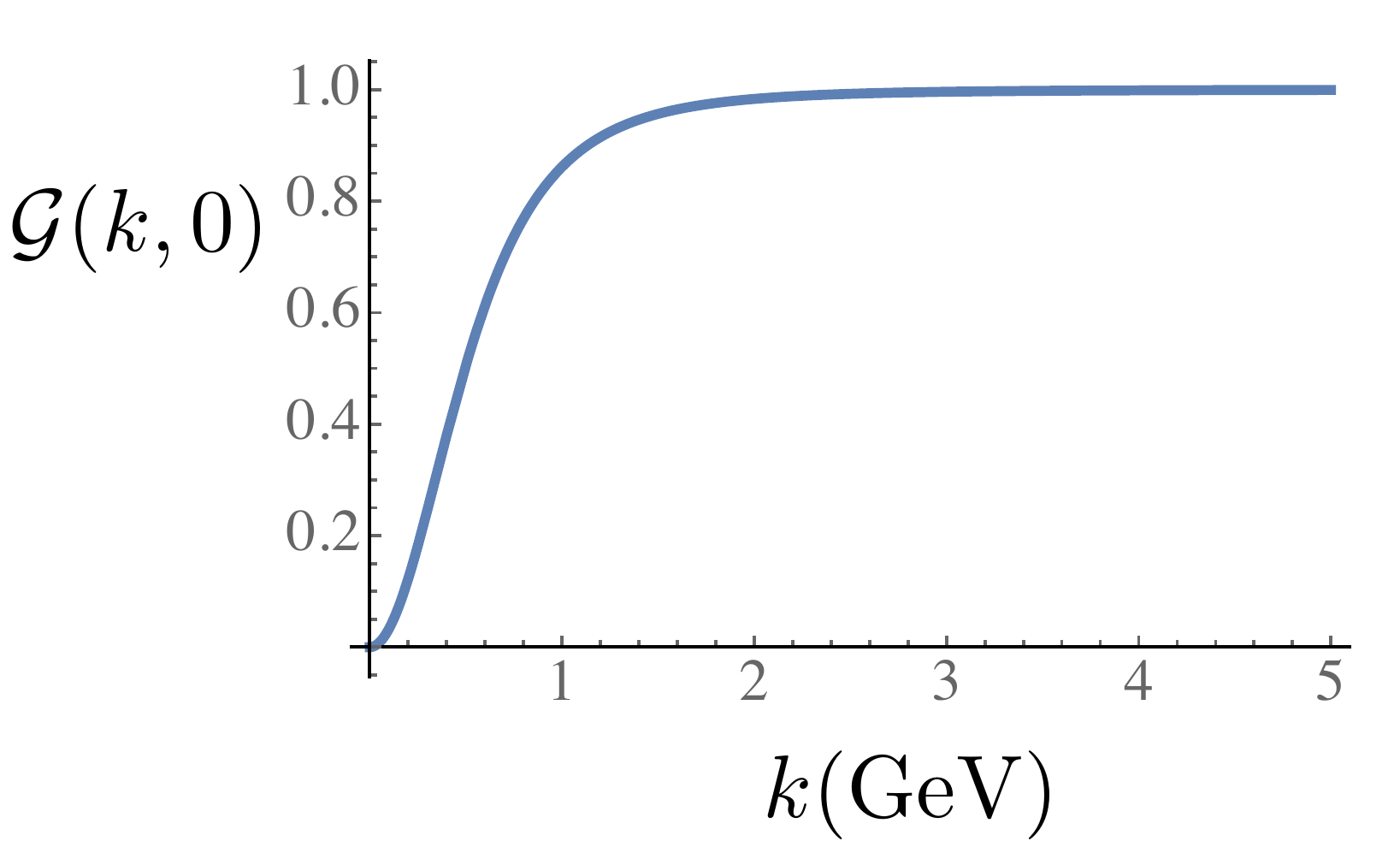}
\caption{$\cG(\vec k,0)$. Cancellation at $k=0$ is needed to prevent infrared divergences from appearing.}
\label{aba:fig1}
\end{figure}

 Space limitations prevent me from saying much more, but it's all included in~\cite{Dumitru:2018vpr}. See also \cite{Dumitru:2019qec} for an interesting application of the formalism.

\section{Summary}
  Ref.~\cite{Dumitru:2018vpr} provides a new way of looking at proton structure that involves 
using  moments of the  color charge density operator. 
Quadratic and cubic correlation functions  in the proton have been constructed for 3-quark light-front  wave functions.  The formalism is general so that evaluations can be made for 
  more complicated wave functions. 
The quadratic correlator ($\r^a\r^b$)   corresponds to Pomeron exchange, and the cubic correlator ($\r^a\r^b\r^c)$ corresponds  to Odderon exchange. 
The present formalism complements the standard GPD formalism.

\section*{Acknowledgments}
I thank  A.~Dumitru  and R.~Venugopalan for many interesting discussions needed to  develope Ref.~\cite{Dumitru:2018vpr}.
I   thank the  MIT-LNS,  the Southgate Fellowship of Adelaide University,, the Bathsheba de Rothchild Fellowship of Hebrew University, the Shaoul Fellowship of Tel Aviv University, the Physics Division of Argonne National Laboratory and the U.S. DOE, Office of Science, Office of Nuclear Physics under Award No. DE-FG02-97ER-41014 for support that enabled this work. 

\end{document}